\documentclass[prb,twocolumn,,showpacs]{revtex4}
\usepackage{graphicx}
\usepackage{hyperref}
\usepackage{amsfonts}
\usepackage{ amssymb}
\usepackage{amsthm}
\usepackage{amsmath}
\begin{document}

\title{Long range Josephson coupling through ferromagnetic graphene}

\author{Ali G. Moghaddam and Malek Zareyan}

\affiliation{Institute for Advanced Studies in Basic Sciences
(IASBS), P.O. Box 45195-1159, Zanjan 45195, Iran}

\begin{abstract}
We study the Josephson effect in graphene-based ballistic
superconductor-ferromagnet-superconductor (SFS) junctions. We find
an oscillatory Josephson coupling $I_c R_N$ of F graphene whose
amplitude is nonvanishing for a half-metallic graphene, increases
for the exchange fields $h$ above the Fermi energy $E_{F}$ and
shows only a slow damping at strong exchange fields $h\gg E_{F}$.
We interpret this long range Josephson coupling as the result of
the exchange mediated Andreev-Klein process at FS interfaces which
enhances the induced antiparallel-spin superconducting
correlations in F graphene by increasing $h$ above $E_{F}$. We
further demonstrate the existence of regular temperature induced
transitions between $0$ and $\pi$ couplings in the plane of $T$
and $h$ where the phase boundaries have distinct shapes at the two
regimes of $h$ below and above $E_{F}$.
\end{abstract}

 \maketitle

Graphene, the two dimensional (2D) solid of carbon atoms with
honeycomb lattice structure, shows unique properties due to its
peculiar gapless semiconducting band structure
\cite{geim-kim,castro-rmp,geim07,katsnelson07}. The conduction and
valence bands in graphene have the conical form at low energies
with the apexes of the cones touching each other at the corners of
the hexagonal first Brillouin zone which determine two
non-equivalent valleys in the band structure. The charge carrier
type (electron-like ($n$) or hole-like ($p$)) and its density can
be tuned by means of electrical gates or doping of underlying
substrate. An important aspect in graphene is the connection
between its specific band structure and the pseudo-spin which
characterizes the relative amplitude of electron wave function in
two distinct trigonal sublattices of the hexagonal structure. This
has caused that the charge carriers in graphene behave as 2D
massless Dirac fermions with a pseudo-relativistic \textit{chiral}
property \cite{geim-kim}. Already anomality of variety of
phenomena including quantum Hall effect \cite{geim-kim}, Andreev
reflection (AR) \cite{beenakker06,beenakker08}, and Josephson
effect \cite{beenakker06-2,heersche} in graphene have been
demonstrated. Here we report on the peculiarity of Josephson
effect in a graphene superconductor-ferromagnet-superconductor
(SFS) junction, which arises from such a Dirac-like spectrum with
chirality. We find that a weakly doped graphene F contact can
support a long ranged \textit{opposite-spin} supercurrent which
persists at strong spin-splitting exchange fields, in striking
contrast to the behavior of the Josephson current in common SFS
junctions \cite{buzdin}.

\par
Superconducting correlations can propagate through a mesoscopic
normal metal (N) contact between two superconductors (S) via the
process of AR at the NS-interfaces in which the subgap electron
and hole excitations with opposite spin directions are converted
to each other \cite{andreev}. Successive AR at the two NS
interfaces and the coherent propagation of the excitations between
these reflections leads to the formation of the so called Andreev
bound states which can carry a supercurrent. The resulting
Josephson effect, characterized by the critical (maximum)
supercurrent $I_c$ and a relation with the phase difference
$\varphi$ between superconducting order parameters of the two
superconductors, is well established in a variety of SNS
structures \cite{likharev}. In an SFS junction, due to the
exchange correlations field $h$, a momentum change of $2h/v_{F}$
between Andreev correlated electron-hole is induced which results
in a damped oscillatory variation of $I_c$ with the length of
F-contact $L$. As the result of the $I_c$-oscillations an SFS
structure can transform into the so-called $\pi$-junction in which
the ground-state phase difference between two superconductors is
$\pi$ instead of $0$ \cite{buzdin,ryazanov}. The damping of $I_c$
occurs over the magnetic coherence length $\xi_h$ which is $\sim
\hbar v_{F}/h$ for a ballistic F \cite{buzdin82,cayssol}. This
makes the Josephson coupling in F junctions rather short ranged as
compared to SNS systems in which the Josephson coupling persists
over much longer lengths of order of the normal-metal coherence
length $\xi_N=\hbar v_{F}/k_{\rm B}T$ \cite{buzdin} (normally $h$
is much larger than the superconducting gap $\Delta$). In
particular for a half-metal F contact with $h\geq E_F$ there is no
Josephson coupling for a sizable contact of length $L\gtrsim
\lambda_F$\cite{cayssol}.
\par
\begin{figure}
\centerline{\includegraphics[width=8cm,angle=0]{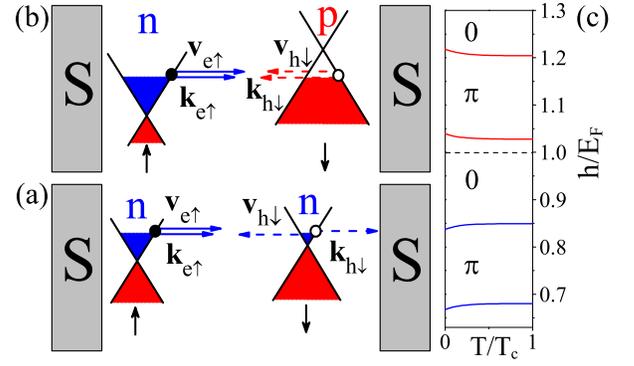}}
\caption{(Color online) (a-b) Schematic of the graphene SFS
junction and the configuration (being electron-like ($n$) and
hole-like ($p$)) of the spin-up and spin-down subbands for two
regimes of (a) $h<E_{F}$ and (b) $h>E_{F}$. The orientation of the
wave vectors and velocity vectors of Andreev correlated
electron-holes (for a normal incidence of electron to the
interface) is also shown. For $h<E_{F}$ the retro reflected hole
has antiparallel momentum and velocity, whereas for $h>E_{F}$
Andreev reflection is specular with the hole having parallel
momentum and velocity. (c) Phase diagram of the $0-\pi$ transition
of the junction of $E_{F}L/\hbar v_{F}=10$ around $h=E_{F}$. The
boundaries between $0$ and $\pi$ phases for $h>E_{\rm}$ and
$h<E_{\rm}$ are the mirror form of each other.} \label{fig1}
\end{figure}

In this work, we demonstrate unusual features of the
exchange-induced $I_c$-oscillations and the corresponding $0-\pi$
transitions in a ballistic F-graphene Josephson contact between
two highly doped superconducting regions (see Fig. \ref{fig1}). We
show that while in the regime of $h<E_F$ the amplitude of the
critical current shows a monotonic damping with the exchange
field, for the higher exchange fields $h\geq E_F$ it develops
drastically different behavior. For a half-metal F with $h=E_F$,
we find that the Josephson coupling $I_cR_N$ ($R_N$ being the
normal state resistance of the junction) has a {\it non-vanishing}
value, in spite of the vanishing density of states for spin-down
electrons. Interestingly, this finite Josephson coupling is
resulted from particular Andreev bound states in which spin-up
propagating excitations and spin-down {\it evanescent} excitations
are involved. These mixed evanescent-propagating states can have
significant contribution in the supercurrent due to the chiral
nature of the carriers in F graphene \cite{beenakker06-2,mz06}.
\par
More surprisingly, we find that for exchange fields above the
Fermi energy $h\gtrsim E_{F}$ the coupling $I_cR_N$ {\it
increases} above its half-metal value and shows damping only at
strong exchange fields $h\gg E_{F}$ with a rate which is much
lower than that of the regime of $h<E_{F}$. We explain this long
range Josephson effect in terms of superconducting correlations
between a $n$-type excitation from the spin-up conduction subband
and a $p$-type excitation from the spin-down valence subband in F
(see Fig. \ref{fig1}b). For $h>E_F$ these two types of excitations
are coupled at the FS-interfaces via a peculiar Andreev process
which is accompanied by a Klein tunneling through the exchange
field $p-n$ barrier \cite{zmg081}. It has been found that this
spin Andreev-Klein process leads to an enhancement of the
amplitude of AR and the resulting subgap conductance of FS
junctions with the exchange field. In the SFS structure the
corresponding Andreev-Klein bound states are responsible for the
long range proximity effect. We further demonstrate the existence
of the temperature-induced regular $0-\pi$ transitions by
presenting phase diagram in $T/T_c$ and $h/E_F$ plane where the
boundaries of $0-\pi$ phases have different forms in two regimes
of $h<E_{F}$ and $h>E_{F}$.
\par
To be specific, we consider a ballistic F-graphene strip of length
$L$ smaller than the superconducting coherence length $\xi=\hbar
v_{F}/\Delta$ which connects two S electrodes (see Fig.
\ref{fig1}). Highly doped superconducting regions can be produced
by depositing superconducting metallic electrodes on top of the
graphene sheet \cite{heersche}. A graphene SFS structure similar
to our setup has been studied by Linder \textit{et al.}
\cite{linder} considering certain values for the Fermi energy in F
and in the electrodes. They concentrated on the existence of a
large residual supercurrent in the points of $0-\pi$ transitions
at $T=0$. Here we consider a more realistic model of highly doped
S electrodes and cover full range of the key parameter $h/E_F$ to
find the above mentioned long range Josephson coupling whose
underlying mechanism will be explained in the following. We take
the Fermi wavelength $\lambda_{F{\rm S}}$ of S-electrodes to be
very smaller than the superconducting coherence length $\xi$ and
the Fermi wavelengths of the two spin subbands in F-graphene
$\lambda_{{F}\sigma}$. By the first condition mean field theory of
superconductivity will be justified and by the second we can
neglect the spatial variation of the superconducting order
parameter $\Delta(x)$ in the superconductors close to the FS
interfaces. Thus $\Delta(x)$ has the constant values
$\Delta\exp(\pm i\varphi/2)$ in the left and right
superconductors, respectively, and vanishes identically in F.
\par
The superconducting correlations between a spin $\sigma$ electron
excitation of wave function $u_{\sigma}$ and the spin
$\bar{\sigma}$ hole excitation of wave function $v_{\bar \sigma}$
can be described by Dirac-Bogoliubov-de Gennes (DBdG)
\cite{beenakker06} equation which, in the presence of an exchange
field, reads:
\begin{equation}
\begin{pmatrix}
\hat{H}_{\sigma}-E_{F}-\varepsilon & \hat{\Delta} \cr {\hat
{\Delta}}^{*} & E_{F}- \hat{H}_{\bar\sigma}-\varepsilon
\end{pmatrix}
\begin{pmatrix}
u_{\sigma}\cr v_{\bar \sigma}
\end{pmatrix}
=0. \label{spinDBdG}
\end{equation}
Here $\hat{H}_{\sigma}=-i\,\hbar
v_{F}(\partial_{x}\hat{\sigma}_{x}+\partial_{y}\hat{\sigma}_{y})-\sigma
h$ and $\hat{\Delta}=\Delta \hat{\sigma}_0$ are the spin-$\sigma$
single-electron Dirac Hamiltonian and the superconducting pair
potential, respectively and $\varepsilon$ is the excitation
energy. The wave functions $u_{\sigma}$ and $v_{\bar\sigma}$ are
two-component spinors of the form $(\psi_{1},\psi_{2})$ and
$\hat{\sigma}_{i}$ ($i=0,x,y,z$) are Pauli matrices, all operating
in the space of two sublattices (pseudo-spin) of the honeycomb
lattice.

\par
Inside F the solutions of DBdG equation (\ref{spinDBdG}) are
electron and hole-like wave functions which are classified by a 2D
wave vector ${\bf k}_{\sigma}\equiv(k_{\sigma},q)$ with the
energy-momentum relation $\varepsilon_{\sigma} =\hbar v_{F}|{\bf
k}_{\sigma}|$. For a finite width $W$ the transverse momentum is
quantized ($q_{n}=(n+1/2)\pi/W$) by imposing the infinite mass
boundary conditions at the edges \cite{berry}. At the Fermi level
$\varepsilon=0$ for a spin direction $\sigma$ and a given $q_{n}$
there are two electron and two hole states with the wave functions
$u_{\sigma}=v_{\sigma}=\exp(\pm ik_{\sigma}x+iqy)(1,\pm\exp(\pm
i\alpha_{\sigma}))$ which are characterized by longitudinal
momentum $k_{\sigma}=\sqrt{k_{{F}\sigma}^2-q^2}$ and the
propagation angle $\alpha_{\sigma}=\arcsin(q/k_{{F}\sigma})$
($k_{{F}\sigma}=(E_{F}+\sigma h)/(\hbar v_{F})$ being the Fermi
wave vector of spin $\sigma$ subband).
\par
The solutions of Eq. (\ref{spinDBdG}) inside S ($h=0$) are rather
mixed electron-hole excitations, the so called Dirac-Bogoliubov
quasiparticles. Assuming ideal FS contacts the electron-hole
conversion can be described by a boundary condition between
electron and hole wave functions which for the left and right
interfaces, respectively, has the forms \cite{beenakker06-2},
\begin{eqnarray}
u_{\sigma}=e^{\mp i\varphi/2+i\beta\mathbf{n}\cdot{\boldsymbol
\sigma}}\,
v_{\bar\sigma},\,\,\,\,\beta=\arccos(\varepsilon/\Delta),\label{e-h}
\end{eqnarray}
where ${\bf n}$ is the unit vector perpendicular to a FS
interface pointing from F to S.
\par
Introducing the normal-state transmission coefficient of spin
$\sigma$ quasiparticles through the junction as
$t_{\sigma}=|t_{\sigma}|\exp(i\eta_{\sigma})=
(\cos\gamma_{\sigma}-i\sin\gamma_{\sigma}/\cos\alpha_{\sigma})^{-1}$
with $\gamma_{\sigma}=k_{\sigma}L$ and imposing the conditions
(\ref{e-h}) at the two FS boundaries ($x =0,L$), we obtain the
following result for the energy of the spin $\sigma$ Andreev bound
state
\begin{eqnarray}
\varepsilon_{\sigma}=\Delta\cos[(\theta(\phi)+\eta_{\bar\sigma}-\eta_{\sigma})/2],
\label{andreev}
\end{eqnarray}
where
$\cos\theta=|t_{\sigma}t_{\bar\sigma}|[\cos\phi+\tan\alpha_{\sigma}\tan\alpha_{\bar
\sigma}\sin\gamma_{\sigma}\sin\gamma_{\bar \sigma}]$.  For a short
junction of $L\ll\xi$ only the Andreev bound states with energies
$|\varepsilon|<\Delta_{0}$ have the main contribution to the
supercurrent. At temperature $T$ the Josephson current can be
obtained from the formula \cite{beenakker91}
\begin{equation}
I=-\frac{2e}{\hbar}\sum_{n,\sigma} \tanh(\varepsilon_{\sigma,
n}/2k_{\rm B}T)\frac{d\varepsilon_{\sigma,n}}{d\varphi},
\label{i-phi}
\end{equation}
where the factor $2$ accounts for the valley degeneracy.
\par
\begin{figure}
\centerline{\includegraphics[width=8cm,angle=0]{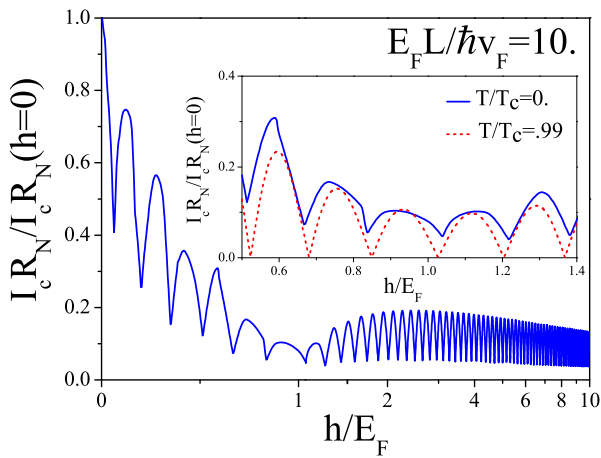}}
\caption{(Color online) Dependence of zero temperature Josephson
coupling $I_{c}R_{N}$ on the exchange energy $h/E_{F}$ scaled with
Fermi energy for the doping $E_{F}L/\hbar v_{F}=10$. Cusp-like
variations indicate $0-\pi$ transitions with a period of order
$hL/\hbar v_{F}$. The coupling has a nonzero value for
$h/E_{F}=1$, develops a smooth maximum for $h/E_{F}\gtrsim 1$ and
shows a slow damping at strong exchange fields.  For $h/E_{F}>1$
the scale of $h/E_{F}$ is logarithmic to clarify the slow decrease
in the Josephson coupling. Inset compares the same dependence at
$T=0$ and $T=0.99 T_c$.} \label{fig2}
\end{figure}
From Eq. (\ref{i-phi}) we have calculated the Josephson critical
current. Figure \ref{fig2} shows the dependence of resulting
Josephson coupling $I_cR_N$ on the exchange energy $h/E_{F}$
scaled in units of Fermi energy at zero temperature and for
$E_{F}L/\hbar v_{F}=10$. Here $R_N=(e^2/\pi\hbar
\sum_{n,\sigma}|t_{\sigma}|^2)^{-1}$ is the resistance of the
corresponding normal (nonsuperconduting) structure. Note that
$R_N$ decreases monotoniacally with the exchange field for
$h>E_F$, due to the linear increase of the density of states in
both spin-subbands. In spite of showing the regular cusps form
variations indicating $0-\pi$ transitions for all $h/E_{F}$, the
overall behavior of the coupling in the regimes of $h<E_{F}$ and
$h>E_{F}$ is drastically different. For $h<E_{F}$ the envelope of
the curve decreases monotonically with $h/E_{F}$ to reach a
minimum for a half-metallic graphene $h=E_{F}$, just similar to
the behavior of $I_cR_N$ in a common SFS junction \cite{cayssol}.
However for $h\geq E_{F}$ the coupling has a finite value and
surprisingly increases smoothly with $h$ before showing a slow
damping at strong exchange fields $h\gg E_{F}$. Thus $I_{c}R_{N}$
as a function of $h$ develops a smooth maximum at an exchange
field which we have found to depend weakly on the doping of F
graphene $E_{\rm}L/\hbar v_{F}$.
\par
We can understand the above behavior of $I_{c}R_{N}$ in terms of
the change in the configuration of the two spin subbands by
varying the ratio $h/E_{F}$. For a normally $n$-doped F graphene
($E_{F}>0$), while for $h<E_{F}$ the charge carriers in both spin
subbands are of the same $n$-type (Fig. \ref{fig1}a), for
$h>E_{F}$  the spin-down carriers turn into the $p$-type as the
Fermi level shifts into the valence subband (Fig. \ref{fig1}b).
For $h<E_{F}$ the Andreev bound states which carry the
supercurrent are made of electrons and holes with the same type
$n$ as shown in Fig. \ref{fig1}a. The AR of the excitations at the
Fermi level is of retro type as in an ordinary FS interface. In
this regime by increasing $h/E_{F}$ the amplitude of AR decreases,
which leads to a decline in the Josephson coupling of the SFS
junction. However when $h>E_{F}$ Andreev correlated electron-hole
pairs in F are of different $n$ and $p$ types which are coupled
via a specular \cite{beenakker06,zmg081} AKR at FS interfaces. Due
to the exchange field induced enhancement of the amplitude of AKR,
the resulted Josephson coupling $I_{c}R_{N}$ shows an increase by
$h$ for $h\gtrsim E_{F}$. But this increase does not continue for
higher $h/E_F$ where $I_{c}R_{N}$ decreases very slowly after
showing a smooth maximum. This slow decline of $I_{c}R_{N}$, in
spite of the increase in the amplitude of AKR, is the result of
superposition of the contributions of different transverse modes
in the supercurrent (see Eq. (\ref{i-phi})).
\par
The behavior of Josephson junction with a half metal graphene
$h=E_{F}$ in which the density of states of the down-spin subband
goes to zero is even more dramatic. In spite of the fact that
there is no propagating excitation at the Fermi level with down
spin, we still find a nonzero critical current for $h=E_{F}$ as it
is seen in Fig. \ref{fig2}. Indeed, in this case we have only
Andreev bound states combined from propagating spin up electrons
(holes) and evanescent spin down holes (electrons). Such Andreev
states exist for all values of $h/E_{F}$, however their
contribution to the supercurrent is negligible unless at the
vicinity of $h=E_{F}$ where they play the main role.
\par
Now let us analyze the effect of a finite temperature. From Eq.
(\ref{i-phi}) we have found that the ballistic graphene SFS
junction can transit from $0$-state to $\pi$-state by varying the
temperature. The resulting phase diagram in the plane of $T/T_c$
and $h/E_{F}$ is shown in Fig. \ref{fig1}c around $h=E_{F}$ and
when $E_{\rm}L/\hbar v_{F}=10$. As it can be seen the values of
exchange fields in which $0-\pi$ transitions occur, increase
(decrease) with temperature for the regime of $h<E_{F}$
($h>E_{F}$) such that the phase boundaries for $h>E_{\rm}$ are the
mirror form of those for $h<E_{\rm}$. This $0-\pi$ phase diagram
is different from that of a common ballistic SFS junction
\cite{chtchelkatchev}. The temperature-induced $0-\pi$ transition
can also be seen from the inset of Fig. \ref{fig2} where we have
compared the oscillations of $I_{c}R_{N}$ at two temperatures
$T=0$ and $T=0.99T_c$. We see that the place of a $0-\pi$ cusp
depends on $T$. This dependence is different in two regimes of
$h<E_{F}$ and $h>E_{\rm}$, which results in different forms of the
phases boundaries as described above. Note that as the result of a
strongly nonsinusoidal current-phase relation at $T=0$, there is a
large residual supercurrent at a transition point \cite{linder}.
As $T\rightarrow T_{c}$ the current-phase relation becomes pure
sinusoidal and the residual supercurrents vanish.
\par
We note that the long range Josephson coupling in graphene SFS
junctions is carried by the superconducting correlations of two
electrons with \textit{opposite} spins. This effect, arising from
the Dirac-like spectrum of excitations and their chiral nature, is
fundamentally distinct from the recently discovered long range
proximity effect in ordinary SFS structures, which was attributed
to the generation of the spin-parallel triplet correlations by
inhomogeneity in the direction of the exchange field
\cite{keizer-bergeret-eschrig}.
\par
The practical importance of the effects predicted here is
connected with the possibility of fabricating high quality FS
structures in graphene which seems to be quite feasible by
considering the recent experimental realizations of proximity
induced superconductivity \cite{heersche,shailos} and
ferromagnetic correlations \cite{vanwees,geim-ieee,vanwees2} in
graphene. In addition to the proximity induced correlations
\cite{brataas}, intrinsic ferromagnetism were also predicted to
exist in graphene sheets \cite{peres} and nanoribbons
\cite{louie}. One alternative way to produce Josephson F contact
would be doping of the spacing part between two S regions (on top
of which metallic superconducting electrodes are deposited) by
magnetic atom impurities \cite{dugaevuchoa08}.

\par
In conclusion, we have demonstrated the existence of a long range
supercurrent in weakly doped graphene ferromagnetic Josephson
junctions. In contrast to the common view, a half metallic
graphene shows a nonvanishing Josephson coupling $I_c R_N$ which
increases by increasing the exchange field $h$ above the Fermi
energy $E_F$, and shows only a slow damping at strong exchanges
$h\gg E_F$. We have explained this long range coupling as the
result of the exchange field mediated Andreev-Klein process at FS
interfaces, which enhances the induction of superconducting
correlations between electrons with opposite spins in F. We have
also presented the $0-\pi$ phase diagram of the coupling in the
plane of $T/T_c$ and $h/E_F$, which reveals the distinct shapes of
the phase boundaries in two cases of $h>E_{F}$ and $h<E_{F}$.

\end{document}